\documentclass[preprint,preprintnumbers,superscriptaddress,amsmath]{revtex4}
\usepackage{url}
\usepackage{epsfig}
\usepackage{graphicx,color}% Include figure files
\usepackage{epstopdf}
\usepackage[psamsfonts]{amssymb}
\usepackage{amsmath}
\usepackage{indentfirst}
\usepackage{ulem}

\begin{document}

\title{Heavy-tailed distribution of cyber-risks}

\author{T. Maillart}
\email{tmaillart@ethz.ch}
\affiliation{Department of Management, Technology and Economics,
ETH Zurich, Kreuzplatz 5, CH-8032 Zurich, Switzerland}

\author{D. Sornette}
\email{dsornette@ethz.ch}
\affiliation{Department of Management, Technology and Economics,
ETH Zurich, Kreuzplatz 5, CH-8032 Zurich, Switzerland}

\begin{abstract}

With the development of the Internet, new kinds of massive epidemics, distributed attacks, virtual conflicts and criminality have emerged. We present a study of some striking
statistical properties of cyber-risks that quantify the distribution and time evolution of information risks on the Internet, to understand their mechanisms, and create opportunities to mitigate, control, predict and insure them at a global scale.
First, we report an exceptionnaly stable power-law tail distribution of personal identity losses per event, ${\rm Pr}({\rm ID~loss} \geq V) \sim 1/V^b$, with $b =0.7 \pm 0.1$.
This result is robust against a surprising strong non-stationary growth of ID losses  culminating in July 2006
followed by a more stationary phase. Moreover, this distribution is identical for different types and sizes of targeted organizations. Since $b<1$, the cumulative number of all losses over all events up to time $t$ increases faster-than-linear with time according to $\mathbf{\simeq t^{1/b}}$, suggesting that privacy, characterized by personal identities, is necessarily becoming more and more insecure.
We also show the existence of a size effect, such that the largest possible ID losses per event grow faster-than-linearly as $\sim S^{1.3}$ with the organization size $S$.  The small value $b \simeq 0.7$ of the power law distribution of ID losses is explained by the interplay
between Zipf's law and the size effect. We also infer that compromised entities exhibit basically the
same probability to incur a small or large loss. 

\end{abstract}

\date{\today}

\maketitle

\section{Introduction}

The Internet has developed into a global system of interconnected computer networks that allows
the exchange of data between millions of private and public, academic, business, and government organizations.
By making possible new forms of social interactions as well as new ways to probe them, 
the Internet provides a unique tool for studying the development and the organization 
of an archetypical  complex system.

But, as in all complex biological and social systems known to us, upgrades of capacity, 
improved networking and additions of functionalities come together with its bundle of parasites,
viruses and criminals. We ask what are the laws, in any, codifying this dynamics, and what are
the possible roles and consequences of such 
apparently negative developments?

In biology, there is a growing realization that evolution has been
driven and shaped by bacteria and viruses \cite{evolvirus}. Similarly, social organizations, 
which are founded on laws 
and regulations, and which are anchored on national (as well as sub- and super-national) boundaries, 
have arguably been shaped in significant part by the need to coordinate and cooperate
in the face of disruptions emerging from within and from the outside. In this vein,
we ask what may the exploding level of criminality and of unlawful exploitation of the Internet
teach us on the organization of other complex systems? Are there robust dynamics or
universal laws that can be inferred and tested? What does the
fact, that electronic crime has appeared and developed concommittantly with the growth
of the Internet, teach us on its organization, its vulnerabilities and its future development?

Given the breadth of these questions, our contribution is to focus on a specific criminality
which is now becoming rampant, the theft of personal information (ID thefts). Using the
most complete dataset from the Open Security Foundation \cite{datalossdb}, we are able 
to identify an explosive growth of ID losses followed by a regime which seems to have
matured into a stationary phase. We document a very heavy-tailed power-law
distribution (an often reported
hallmark of complex systems) of severities of ID theft events, which is robust over all time periods
and across different types of social organizations (private and public). By quantifying
the scaling of losses as a function of organization sizes, we unearth
a significant size effect.

\section{Maturation and severity of ID losses:  non-stationary and stationary properties}

\subsection{Contextual data description}

 From early (gentle) hackers breaking in systems to demonstrate  their skills, some turned into 
seasoned ``black hats'' making money as part of an explosively growing business based on 
ubiquitous Internet insecurity\cite{frei2006, zittrain2008}. Compared with the attacks that used to disrupt 
network on a large scale, most electronic attacks nowadays 
extract out valuable data while remaining quite furtive \cite{mansell2005tac}. 
This can be likened to an electronic form of massive parasitism.
In terms of monetary value and volume, one of the largest types of data targeted by pirates 
is personal identity information (ID), such as credit card numbers, social security numbers, 
banking accounts, and medical files.
Since each ID theft or leakage is a ``loss of control'' of one's individual private data, it can be considered already as a damaging event, forerunning the potential realized financial and/or social losses \cite{kanderson2008}. Actually, stealing ID's is the goal which is common to a wide spectrum of non-destructive Internet attacks focused on profit,  from botnets to highly tailored attacks \cite{dagon2007,botnet_idtheft_case,schneier2005rtp,koops2006iti}. 
The (uncontrolled) dissemination of personal information raises the important social issue of people's identity resilience in the information technology era \cite{kanderson2008,mansell2005tac}.
In our quantitative study of cyber-risks, we take a
ID theft as a usable elementary unit of cyber-risks, for two main reasons. First, it provides a natural metric of the 
``permeability'' of information systems, guiding towards the identification of the underlying mechanisms. Second, 
it offers a common basis, or currency, to compare a large variety of heterogeneous events involving
many different types of organizations. 

ID loss event data have been thoroughly collected by several independant organizations. We use the most complete dataset from the Open Security Foundation \cite{datalossdb}, that contains 956 documented events reported mainly in the USA between year 2000 and November 2008. 
The catalog provides also the involved organization, the date and amount of loss (measured as the numbers of ID stolen). Data are homogeneously sampled among various types of organizations:  business (35\%), education (30\%), governments (24\%) and medical institutions (10\%). We define an event following the procedure
described in Ref.\cite{datalossdb,hasan2006sad}. For instance, the largest entries in the data set are (i) the discovery and disclosure of an attack over several years of the TJX Companies with a probable exposition of more than 90 millions IDs (end of the event: January 2007), (ii) the Cardsystems' hack impacting 40 million Visa, MasterCard and American Express cardholders (June 2005), (iii) America Online (30 million credit card ID exposed in 2004), and (iv) the U.S. Department of Veterans Affair (more than 25 million of ID stolen in 2006).  While there is  not warranty
of completeness, our tests below suggest that the catalog of the Open Security Foundation provides
a reasonable representative sample of the overall activity of ID thefts occurring on the Internet.

\subsection{Transition from explosive growth to statistical stationarity}

\begin{figure}
\centerline{\epsfig{figure=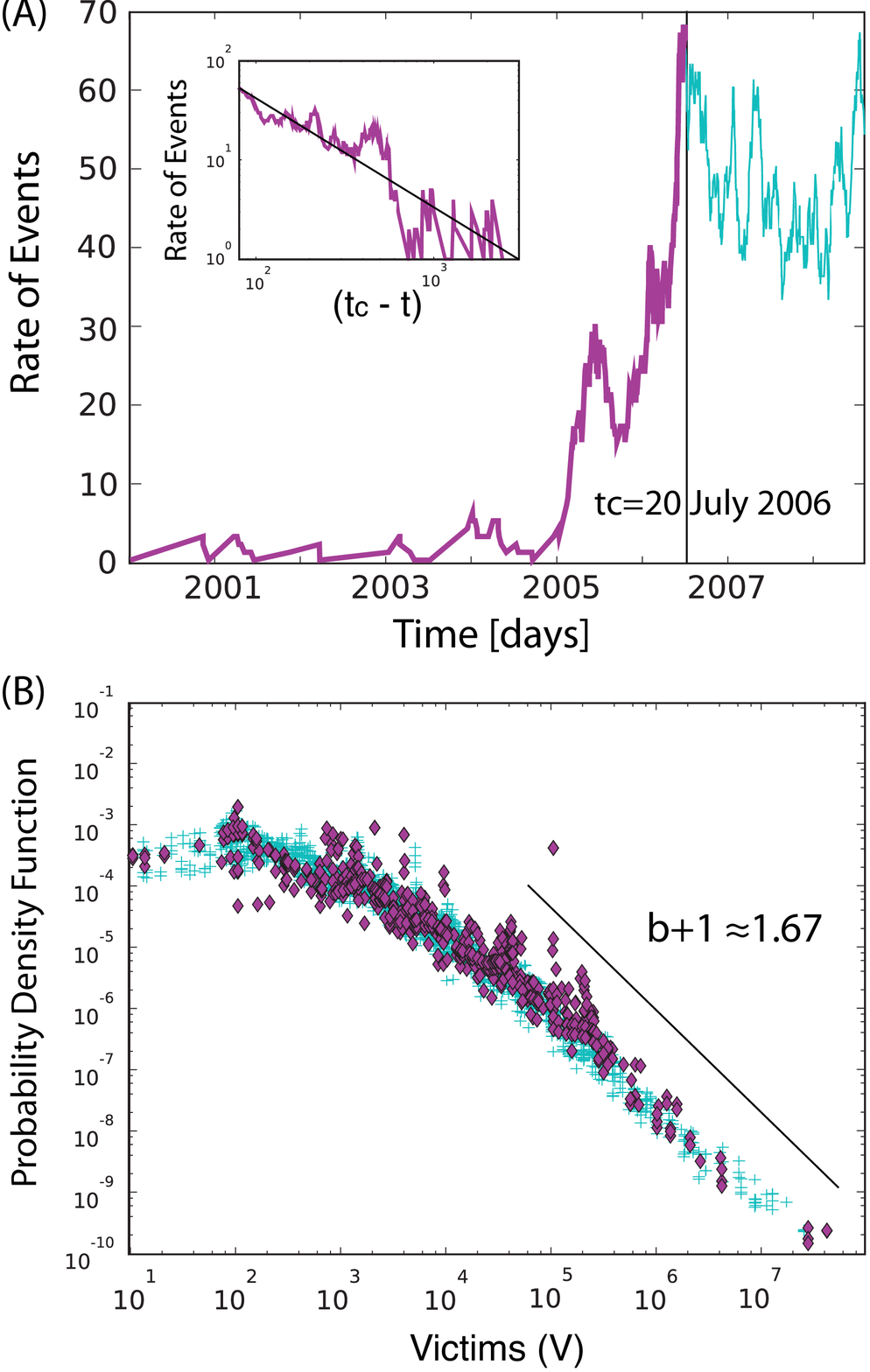,angle=0,width=8cm,scale=1}}
\caption{(colors online) (A) The rate of ID loss events in sliding windows of fifty days is plotted as a function of time,
revealing the existence of two successive regimes: (i) explosive growth culminating in July 2006 (red thick line) and (ii) stable rate thereafter (blue thin line). The inset shows the logarithm of the rate of ID loss events as a function of $(t_{c}-t)$ with $t_{c}= 20/07/2006$, 
such that a straight line qualifies a super-exponential singular acceleration $\sim 1/(t_{c}-t)^m$ with $m \approx 1$.
(B) Scatter proxies of probability density functions (PDF) of the size of events obtained in sliding windows of $100$ days duration. PDFs obtained by binning or with the adaptive Gaussian kernel density estimator \cite{cranmer2001} provide similar results. The size of an event is defined as the total number of IDs lost in that event. For the sake of clarity, we show only one PDF out of every fifty PDFs. Red diamonds (respectively blue crosses) correspond to the PDFs obtained before (respectively after) the peak in July 2006.}
\label{fig:sliding_win}
\end{figure}

The total rate $C(t)$ of ID theft events (measured by the
number of events in a sliding window of 50 days) is shown in the top panel of Figure \ref{fig:sliding_win} 
as a function of time. This panel reveals the existence of two distinct phases. 
Starting from 2000, one can observe a dramatic increase of the rate
of attacks up to a peak reached in July 2006, followed by a plateau thereafter. 
The inset shows a simple non-parametric test suggesting that the first regime was
characterized by a faster-than-exponential growth. Such singular behavior
characterized by a transient explosive growth mathematically modeled
by a power law finite-time singularity is the diagnostic of an impending change of regime
\cite{JS2001,IS2002,GS2002},
which we indeed observe beyond the peak in July 2006. This suggests to interpret
the time evolution of the rate of ID loss events 
as first undergoing a non-sustainable growth followed by a maturity period 
which characterizes the present epoch.

The lower panel of Figure \ref{fig:sliding_win} demonstrates that
the distribution ${\rm pdf}(V)$ of event sizes (defined as
the total number of ID stolen in that event) has remained stable, within statistical fluctuations,
over the whole time period investigated here from 2000 to Nov. 2008. 
There is no significant difference between the probability density functions (PDF) in the growth regime
before July 2006 (red circles) and during the maturity period (blue diamonds),
as evidenced by the perfect collapse of the PDFs.  Indeed, Q-Q plots of one sample
as a function of other samples and in function of the entire sample, were found to be approximately linear with slope ${\rm slope} \approx0.9\pm0.3$.
This suggests that
the mechanism underlying the loss of ID has remained stable, notwithstanding the enormous evolutions that have occurred over this whole time period.

The two pieces of information provided by the two panels of Figure \ref{fig:sliding_win} imply that
the rate $N(V,t)$ of events of size $V$ occurring at time $t$ can be factorized under the form
\begin{equation}
N(V,t)=C(t)\cdot {\rm pdf}(V)\ ,
\label{sfdk}
\end{equation}
where $C(t)$ and ${\rm pdf}(V)$ constitute two
independent contributors to cyber-risks. The macro-variable $C(t)$ embodies
 the overall evolution of the level of threat associated with ID losses. In other words, it provides
a metric quantifying the systemic ``state of insecurity'' of the Internet. In contrast, ${\rm pdf}(V)$
measures the relative frequency of large versus small ID losses. While the rate of attacks
has varied enormously between 2000 and 2008 as shown by the behavior of $C(t)$ in the
upper panel of Figure \ref{fig:sliding_win}, the relative frequencies of various event sizes
has remained remarkably stable, as shown in the lower panel of Figure \ref{fig:sliding_win}.
We now turn to the determination of ${\rm pdf}(V)$ in order to characterize
quantitatively the level of cyber risks per event.

\section{Distribution of ID theft event sizes and consequences}
 
\subsection{Power-Law versus Stretched Exponential}

\begin{figure}
\centerline{\epsfig{figure=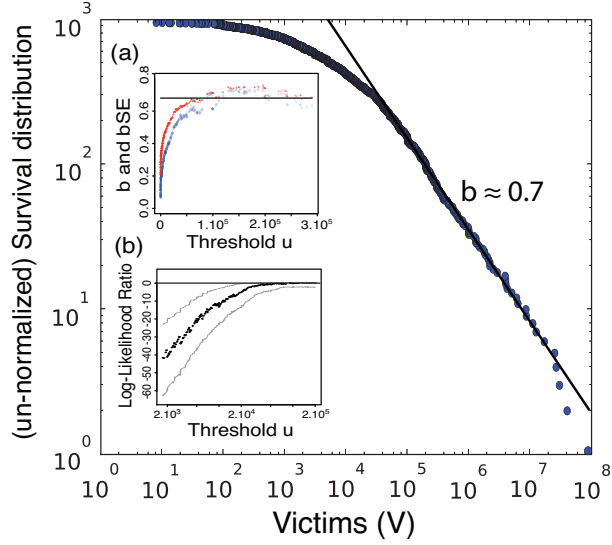,angle=0,width=8cm, scale=1}}
\caption{(colors online) Non-normalized survival distribution (double logarithmic scale) of ID losses, constructed using the data provided in \cite{datalossdb} The straight black line is the fit with the power law (\ref{hynhte}) with $b =  0.7$ for number of victims larger that the lower threshold  $u = 7 \cdot 10^4$. The red dashed line is the fit with the Stretched Exponential (SE) defined by expression (\ref{sekjgktr}). Inset (A) shows the dependence of the index $b$ as a function of $u$ obtained directly from the maximum likelihood estimation (MLE) of the exponent of the power law  (\ref{hynhte}) (crosses) and indirectly from the MLE of the parameters $c,d$ of the stretched exponential (SE) law (\ref{sekjgktr}) using the correspondence $b_{\rm SE} =  c (u/d)^c$ (diamonds) as described in the text. The horizontal line is at $b=0.68$. Inset (B) shows the logarithm of the likelihood ratio (LLR) of the power law versus the SE fits, which converges to $0$ as $u$ increases, thus demonstrating that the simple one-parameter power law is sufficient and the two-parameter SE law is not necessary to explain the tail of the data set. The two grey lines delineate the 95\% confidence interval obtained by bootstrap.}
\label{fig:idtheftfit}
\end{figure}

Given the result of the previous section that a unique distribution ${\rm pdf}(V)$ is
sufficient to describe the frequency of event sizes in all time windows from 2000 to 2008,
we now determine ${\rm pdf}(V)$ by using the largest
possible statistical sample including all events of this period.
Figure \ref{fig:idtheftfit} presents the (non-normalized) empirical survival (also called complementary cumulative) distribution function ${\bar F}_{u}(V)$, defined as the probability that the number of victims in a given event is larger than or  equal to $V$ in the range $V \geq u$. Note that ${\bar F}_{u}(V)$ has a shape similar to the PDFs shown in the lower panel of Figure \ref{fig:sliding_win} with an approximately straight tail in this double-logarithmic scale, 
suggesting a power law distribution
\begin{equation}
{\bar F}_u(V)=\left({u \over V}\right)^b~,\ \ {\rm for }\ \ V \geq u\ .
\label{hynhte}
\end{equation}
This power law (\ref{hynhte}) is observed over more than three decades 
above the lower threshold $u\approx 7.10^4$.
A maximum likelihood estimation (MLE) of the exponent determines $b=0.7 \pm 0.1$.
If model (\ref{hynhte}) is a correct description of
the survival distribution, then ${\rm pdf}(V) \sim 1/V^{1+b}$, which is shown as a straight line with slope $-1.7$
in the lower panel of Figure \ref{fig:sliding_win}.
This result suggests that ID thefts have statistics similar to those observed in
the large class of systems with heavy-tails, such as firm and city sizes
in the social sciences or earthquakes and other calamities in the natural sciences.

However, visual evidence and MLE are not sufficient to demonstrate that 
the power law (\ref{hynhte}) is adequate to describe our statistical data of ID thefts,
as discussed in several earlier works \cite{LS1998,malevergne2005eds,ClausetNewman2007}.
To prove that the one-parameter power law (\ref{hynhte}) is sufficient, we embed it into a
broader two-parameter law that have previously been reported to provide a flexible 
model of many empirical fat-tailed distribution \cite{LS1998}
and perform a standard log-likelihood ratio (Wilks) test.
Specifically, we use the ``stretched exponential'' (SE) family
\begin{equation}
 {\bar F}_{u}(V)=\exp \left[-\left({\frac{V}{d}^c}\right)+ \left({\frac{u}{d}^c}\right) \right] \ ,\ \ \ {\rm for}\ V \geq u\ ,
\label{sekjgktr}
\end{equation}
where $c$ and $d$ are respectively the shape and scale parameters of the SE distribution.
Malevergne et al. \cite {malevergne2005eds} have shown that the power law family (\ref{hynhte}) is asymptotically
embedded in this SE family in the limit
\begin{equation}
c\cdot \left({\frac{u}{d}}\right)^c \rightarrow b,\  {\rm as}\ \ c \rightarrow 0\ . 
\label{grkfnw}
\end{equation}
This has two practical applications: (i) the calibration of $c$ and $d$ for a given $u$ provides
an alternative determination (using (\ref{grkfnw}) of the exponent $b$ of the power law (\ref{hynhte}) if $c$ is indeed small (typically less than $0.3$);
(ii) we can use the formal likelihood ratio test of embedded hypotheses which has been
shown to hold for the power law seen as asymptotically embedded in the SE family  \cite{malevergne2005eds,malevergne2006efr}, to determine whether the one-parameter power law is sufficient or a two-parameter distribution like the SE is necessary. 
Inset (a) in Fig \ref{fig:idtheftfit} shows the estimated exponent $b$ obtained from the maximum likelihood
estimation (MLE) of $c$ and $d$ translated into $b$ via the equation
$b_{\rm SE} =  c (u/d)^c$ derived from (\ref{grkfnw}), as a function of the lower threshold $u$. For $u \geq 7 \cdot 10^4$,
we obtain an excellent confirmation of the value $b \simeq 0.7 \pm 0.1$ determined from the direct MLE of the power law (\ref{hynhte}).
Inset (b) in Figure \ref{fig:idtheftfit} shows in addition the logarithm of the likelihood ratio (LLR)
of the power law versus the SE fits: for $u < 7 \cdot 10^4$, LLR$<0$ indicating that the power law
is not sufficient and that the SE is necessary; in contrast, for $u \geq 7 \cdot 10^4$, the power law
is sufficient and the SE is not necessary, degenerating into the power law as the condition
(\ref{grkfnw}) becomes valid.

\subsection{Evidence for incompleteness of reported losses for small event sizes}

We now discuss two possible hypotheses for  
the observed cross-over at $u \approx 7 \cdot 10^4$ below which the distributions shown
in the lower panel of Figure \ref{fig:sliding_win} and in Figure \ref{fig:idtheftfit}
exhibit a significant downward curvature characterizing a deviation from the power law
(\ref{hynhte}). 

A first possible interpretation is that this deviation from the power law 
reflects the fact that hackers are preferentially targeting
large organizations offering substantial potential gains. As a consequence, there 
would be practically no ID thefts involving only a few individuals. This hypothesis does not stand
closer scrutiny: there is strong evidence that millions of home computers are compromised  \cite{botnet_idtheft_case}
via the use of botnet deployment mechanisms centrally managed by pirates \cite{dagon2007}, with
each computer infection being a unique event potentially leading to ID thefts limited to those IDs which are
stored in the computer. 
According to Vinton Cerf's, $100-150$ millions computers over a total of $600$ millions are compromised\cite{cerf2008}. As a rough estimation, assuming that all computers
have about the same probability of being infected and counting one computer
per Internet user, this implies that about 
one sixth of US computers are exposed. Thus, about 50 millions US citizen are constantly exposed to attacks targeting their own computer.
Such events should thus provide a huge population of small ID theft events' which is absent 
from even the most complete dataset of the Open Security Foundation \cite{datalossdb}.

\subsection{Super-linear growth of the ID loss threat}

There is another remarkable consequence deriving straightforwardly from
the power law (\ref{hynhte}) with exponent $b<1$.
Indeed, the smallness of the power law exponent $b<1$ implies a typical faster-than-linear growth 
of cumulative losses with time. Because $b<1$ and assuming that there are 
no upper threshold yet relevant, the mean and variance of the number of ID losses per event are mathematically infinite.
In practice, this means that their values in any finite catalog exhibit growing random fluctuations as the number of recorded events increases, due to the never decreasing influence of the largest event sizes. Then, the cumulative sum ${\cal V}(t)$ of all losses over all events up to time $t$
is controlled by the few largest events in the catalog \cite{sornette2006cpn}. This leads
to a faster-than-linear growth 
\begin{equation}
{\cal V}(t) \sim t^{1/b} \approx t^{1.4}~.
\label{tjhwbwda}
\end{equation}
This results is
solely due to the statistical mechanism that, as more events occur, 
some are bound to explore more and more the tail of the 
heavy-tailed power law distribution (\ref{hynhte}). Note this law (\ref{tjhwbwda})
constitutes a lower bound, which is attained only when the rate of 
event occurrences is itself not growing, as seems to be the 
case since July 2006.

Such faster-than-linear growths due to 
the pure statistical power law mechanism have been documented in natural hazards for losses caused by floods \cite{pisarenko1998nlg} 
and for the cumulative seismic energy released at regional scales \cite{rodkin} (see \cite{sornette2006cpn}  for a detailed mathematical derivation and discussion).
Given the heavy-tail nature of the distribution of ID theft numbers per event, we should  not be surprised that the Internet appears more and more insecure and dangerous, just as a result of this mechanism.

\section{In cyber-risks, size matters}

\subsection{Cross-sectional universality of ID losses}

We have shown that the PDF of event sizes is constant over time.  We now
investigate whether there exists some difference between the
PDFs of event sizes in a cross-sectional analysis of different sectors of activity, which could
reveal different vulnerability characteristics. 

\begin{figure}
\centerline{\epsfig{figure=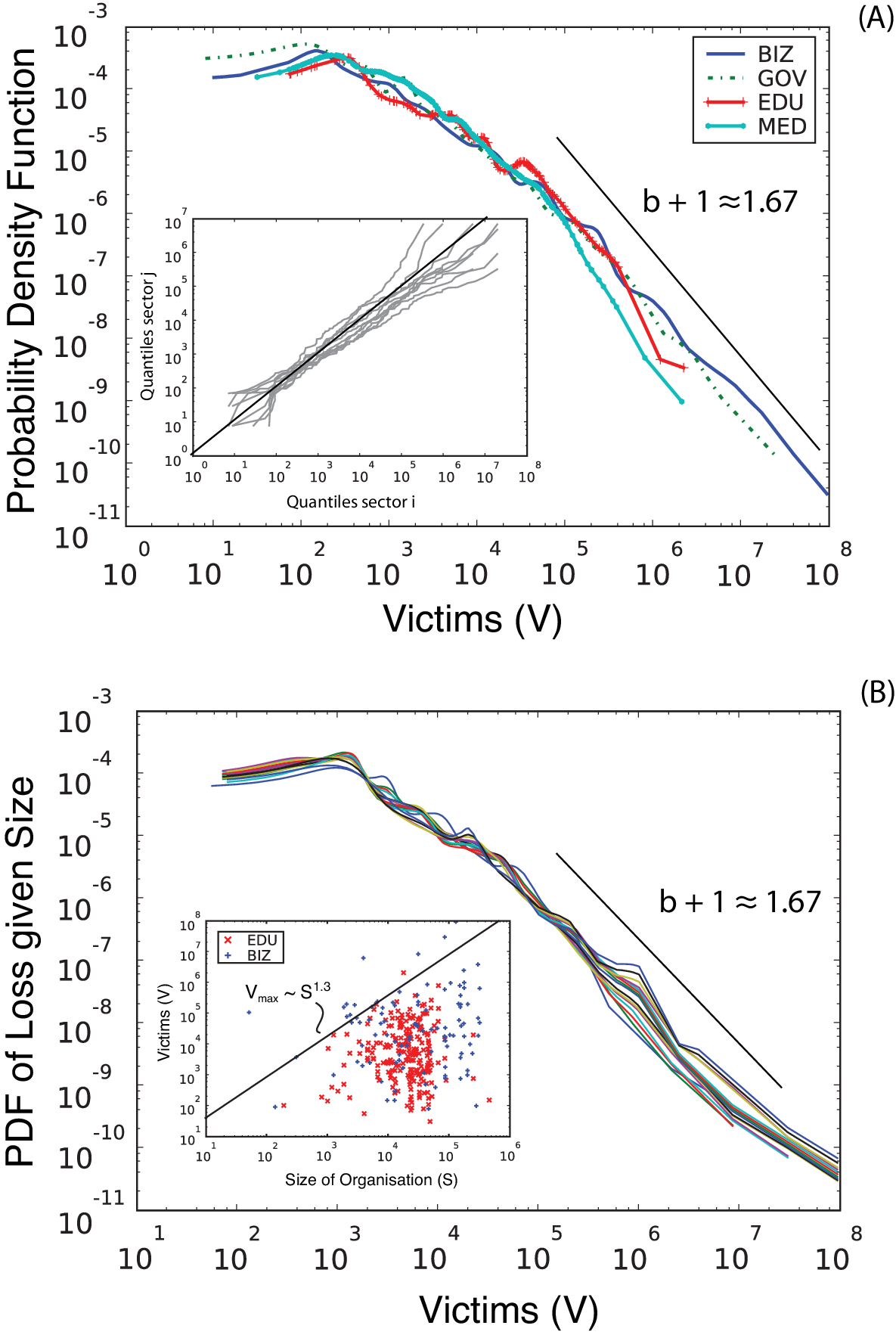,angle=0,width=7cm, scale=1}}
\caption{(colors online) (A) Probabibility density functions of the number of victims ($V$) per event
sorted by sector: business (Biz), governmental agencies (Gov), schools and universities (Edu), medical industries (Med). Inset shows quantile-quantile plot (with 5\% interquantiles) of sectors taken against each other. Linear fit obtained for the presented lines show that we cannot reject that $slope=1$ , ruling out  the hypothesis that distributions are different.
(B) Probability density functions (PDF) of victims per event sorted by sizes of the target organizations. We construct one PDF per decade in organization sizes, i.e., we collect all events occurring for organizations of sizes between $S^*$ and $10 \times S^*$ and construct the corresponding PDF. We then vary $S^*$ across the whole sample (to avoid overlapping we take only one out of fifty PDFs). All PDFs exhibit a good collapse, confirming the universality of the power law distribution of event loss sizes, as in
Fig.\ref{fig:sliding_win} and Fig. \ref{fig:idtheftfit}. Similarly to presented above, by performing linear regressions of (log) quantiles of all samples, we cannot rule out that all samples are drawn from the same probability distribution. The inset shows in double logarithmic scale a scatter plot of the losses ($V$) as a function of size for 374 entities. The straight line with slope $\simeq 1.3$ is the best linear fit ($p=0.00$ and $R^2=0.74$) of the $99\%$ percentile of the logarithmic losses for both 269 universities (blue plus symbols)\cite{univ} and 105 publicly traded companies (red crosses)\cite{bloomberg} as a function of organization logarithmic size.}
\label{fig:loss_size_slide}
\end{figure}

Our datasource uses four distinct sectors of activity:  publicly traded companies (Biz), schools and universities (Edu), governmental agencies (Gov), and medical services (Med). 
Distinct regulations and industry benchmarking imply that 
organizations implement homogenous security processes in a given sector, but 
these security processes operating in a given sector are different
from those in a different sector.
A priori, one could expect that distinct factors acting in these different
sectors imply dissimilar attractiveness to hackers leading to different levels of vulnerability,
which should be revealed in the statistical properties of the catalogs of ID losses. In contradiction
with this anticipation, the top panel of
Figure \ref{fig:loss_size_slide} shows that one cannot reject the hypothesis
that the PDFs of ID loss size per event are identical for the 
four sectors Biz, Edu, Gov, Med. 

If two typical organizations belonging 
to two different sectors are subjected to distinct exposition and permeability threats,
the remarkable conclusion suggested by the top panel of Figure \ref{fig:loss_size_slide} is that
the associated level of security just compensates for the
increasing threat, putting all organizations at a similar overall risk level.
This result is reminiscent of the effect documented in Refs.\cite{carlson2000PRL,doyle2005pnas},
that systems exposed to different distributions of attacks converge to similar
level of vulnerabilities when they try to optimize their efficiency in the presence of
constraints. This could mean that organizations, which are indeed attempting to 
optimize their defenses against cyber-risks, may have already reached
an intrinsic barrier. With the 
evolving nature of the threats and given the complexity of the associated processes
in the presence of limited resources, the observed level of ID losses may be
a robust dynamical equilibrium that will be difficult to improve upon. This suggests
that, in absence of a fundamentally new qualitative paradigm,
these cyber-risks are bound to remain with us for the foreseeable future.

\subsection{Size effects of vulnerabilities to cyber-risks}

The bottom panel of  Figure \ref{fig:loss_size_slide} plots the PDFs of
victims per event sorted by target organization sizes. There are several
possible measures for the size of an organization. Here, we take  
the number of employees, which correlated well with other measures \cite{axtell2001zdu}. 
The PDFs are constructed for 269 universities \cite{univ} and 105 publicly traded companies \cite{bloomberg}.
The good collapse of the PDFs confirms the universality of the power law distribution of event loss sizes, as in
Fig.\ref{fig:sliding_win} and Fig. \ref{fig:idtheftfit}. 

However, the tails of the PDFs are truncated at upper
values which seem to grow with the organization sizes. This size effect is better revealed
by the scatter plot of the inset in the bottom panel of  Figure \ref{fig:loss_size_slide}, which shows that
the largest losses $V_{\rm max}$ for a given range of organization sizes $S$ seem to grow with $S$.
This visual impression is confirmed by performing linear regressions of 
$\log V(q)$ as a function of $\log S$, $\log V(q)= \sigma \log S + \epsilon$, where $V(q)$ is the  
99\% quantile of the losses for a given organization size $S$. 
We find a stable determination of the exponent $\sigma \approx 1.3 \pm 0.1$. This means that the
largest losses for a given set of entities of size $S$ grow with $S$ as $V_{\rm max} \sim S^\sigma 
\approx S^{1.3}$. 

Naively, one would have expected a linear growth with $\sigma=1$. The faster-than-linear law
may express a combination of effects, which include a faster-than-linear growth of the 
number of IDs stored in a given entity as a function of its number of employees, a 
bigger exposition that makes the attacks of large entities more attractive to hackers
and possibly a greater vulnerability due to more bridges or ``boundaries'' with the external world 
which are more difficult to manage. The faster-than-linear law is characteristic of a size effect
which is similar to the size effects 
documented for instance in material failure  \cite{bavzant1997sqf} and species fragility \cite{cardillo2003bde}.

We now show how $\sigma$ is related to the exponent $b$ of the PDFs of event loss sizes defined in 
(\ref{hynhte}). For this, we write the probability ${\rm Pr}({\rm ID~ losses} \geq V)$ to find an event with more than $V$ ID losses as
\begin{equation}
{\rm Pr}({\rm ID~ losses} \geq V) = \int_{S_{\rm min}}^{+\infty} dS \cdot Z(S) \cdot {\rm Pr}_1({\rm ID~ losses} \geq V | S)~,
\label{htbalaaa}
\end{equation}
where $S_{\rm min}$ is a minimum size for  an organization to be viable, and $Z(S)$ is the distribution of organization sizes, well-known to follow Zipf's law ($Z(S) \sim 1/S^{1+\mu}$ with $\mu \approx 1$) \cite{axtell2001zdu,zipf1949hba,gabaix1999zsl} so that $Z(S) \cdot dS$ is the number of organizations with sizes between $S$ and $S+dS$. Moreover,  ${\rm Pr}_1({\rm ID~ losses} \geq V | S)$ is the probability to find an event with more than $V$ ID losses in a given organization of size $S$. 
We know one property of ${\rm Pr}_1({\rm ID~ losses} \geq V | S)$, namely that it drops abruptly to vanishing values for
$V > C \cdot S^\sigma$, where $C$ is a positive constant, as documented above. This implies that, for a fixed $V$,
all integrants with $S < (V/C)^{1/\sigma}$ do not contribute to the integral. Motivated by the power law (\ref{hynhte}),
we also assume a power law shape for ${\rm Pr}_1({\rm ID~ losses} \geq V | S)$ with exponent $b_1$. Putting all
this together, expression (\ref{htbalaaa}) becomes
\begin{equation}
{\rm Pr}({\rm ID~ losses} \geq V) \simeq  \int_{S_{\rm min}(V)}^{+\infty} {dS \over S^{1+\mu}} \cdot {1 \over S^{b_1}}~,
\label{affyhjkik57k}
\end{equation}
with $S_{\rm min}(V) \sim (V/C)^{1/\sigma}$.
This yields ${\rm Pr}({\rm ID~ losses} \geq V) \sim 1/S^{b_1 +{1 \over \sigma}+(\mu-1)}$. Identifying this power law 
with (\ref{hynhte}) in the tail gives $b = b_1 + {1 \over \sigma}+(\mu-1)$. Given that $\sigma \approx 1.3 \pm 0.1$,
we have $1/\sigma \approx 0.77 \pm 0.1$. Since $b=0.7 \pm 0.1$, 
this calculation allows us to infer that the distribution of ID losses for a given organization
is fairly flat ($b_1 \simeq 0$). In other words, the efforts
necessary to get just a few or a large number of IDs are not much different, 
once an organization has been compromised. Our conclusion does not rely
sensitively on the validity of Zipf's law. However, the value $b<1$ imposes a bound
on the exponent $\mu$ of Zipf's law which cannot be significantly larger than $1$.

\section{Conclusion}

We have presented three different tests that confirm the general validity and robustness of the probability distribution of ID losses per event (where ID losses has been taken as a proxy for information risks on the Internet). We showed that the PDFs are the same irrespective of (i) the growth phase before July 2006 versus stationary regime thereafter, (ii) the 
sectors of activity, and (iii) the size of targeted organisations. Returning to the questions raised in the introduction, it is striking and a priori counter intuitive to find that all organisations are evenly vulnerable, whatever their implemented information security.  This raises important questions concerning the tradeoff between exposition and counter-measures in the complex evolving landscape of cyber-risks. The consequences on the evolution of the Internet remain to be studied. This present paper provides a first partial approach of the study of the development of the Internet and of cyber-risks taking into account their intricate entanglement.

We have shown the existence of a size effect, such that the largest possible ID losses per event grow faster-than-linearly with the organization size. This has led us to derive two important consequences. First, the small value $b \simeq 0.7$ of the power law distribution of ID thefts is explained by interplay
between Zipf's law and the size effect. Second, we have found indirect evidence that compromised entities
typically expose to hackers a small or large number of IDs with basically the same frequency. 
This inference is very important for the quantification of cyber risks and suggests that
counter-measures should be targeted towards building internal barriers, avoiding the 
``Titanic'' effect of inadequate compartmentalization.

\vskip 1cm
{\bf Acknowledgements}:  This work was supported by  the Swiss National Foundation, grant 2-77059-07 and by the MTEC Foundation (Fordergesellschaft fur Betriebswissenschaften MTEC), We also acknowledge financial support 
from the ETH Competence Center ``Coping with Crises in Complex 
Socio-Economic Systems'' (CCSS) through ETH Research  Grant CH1-01-08-2.

\newpage
%----------------------F-I-G-U-R-E-S------------------------

%FIGURE 1

%FIGURE 2

%FIGURE 3


\begin{thebibliography}{}
\bibitem{univ}
Source: Official websites from universities. University population is composed with students, graduates, administrative staff and faculty.
\end{thebibliography}
\end{document}